# Sensitivity of micromechanical actuation on amorphous to crystalline phase transformations under the influence of Casimir forces

M. Sedighi, W. H. Broer, G. Palasantzas*, and B. J. Kooi

Zernike Institute for Advanced Materials, University of Groningen, Nijenborgh 4, 9747 AG Groningen, The Netherlands

**Abstract**

Amorphous to crystalline phase transitions in phase change materials (PCM) can have strong influence on the actuation of microelectromechanical systems under the influence of Casimir forces. Indeed, the bifurcation curves of the stationary equilibrium points and the corresponding phase portraits of the actuation dynamics between gold and AIST ($Ag_5In_5Sb_{60}Te_{30}$) PCM, where an increase of the Casimir force of up ~25% has been measured upon crystallization, show strong sensitivity to changes of the Casimir force as the stiffness of the actuating component decreases and/or the effective interaction area of the Casimir force increases, which can also lead to stiction. However, introduction of intrinsic energy dissipation (associated with a finite quality factor of the actuating system) can prevent stiction by driving the system to attenuated motion towards stable equilibrium depending on the PCM state and the system quality factor.

______________________________________
*Corresponding author: g.palasantzas@rug.nl

# I. Introduction

Nowadays fluctuation induced electromagnetic (EM) forces between neutral bodies become increasingly important in microelectromechanical systems (MEMS) [1]. These forces between two objects arise due to perturbation of quantum fluctuations of the EM field [1-12], as it was predicted by H. Casimir in 1948 [2] assuming two perfectly conducting parallel plates. Following Casimir's calculation, Lifshitz and co-workers in the 50's [3] considered the general case of real dielectric plates by exploiting the fluctuation-dissipation theorem, which relates the dissipative properties of the plates (optical absorption by many microscopic dipoles) and the resulting EM fluctuations. The theory correctly describes the attractive interaction due to quantum fluctuations for all separations covering both the Casimir (long-range) and van der Waals (short-range) regimes [1, 3].

The dependence of the Casimir force on materials is an important topic since in principle one can tailor the force by engineering the boundary conditions of the electromagnetic field with a suitable choice of materials [5-12]. The latter allows the exploration of new concepts in actuation dynamics of MEMS. This is because MEM engineering is conducted at the micron to nanometer length scales. As a result Casimir forces become of increasing interest [1] because MEMS have surface areas large enough but gaps small enough for the Casimir force not only to draw components together but also to lock them permanently [1, 4, 13-25]. This effect is known as stiction causing device malfunction. On the other hand, the irreversible adhesion of moving parts resulting in general from Casimir and electrostatic forces can be exploited to add new functionalities to MEMS architectures [1]. Therefore, Casimir interactions will inevitably need to be faced with particular attention to the troublesome pull-in instabilities for example in micro switches [1, 13-25].

In fact, micro switches are essential MEMS components that are typically constructed from two electrodes of which one is fixed, and the other is suspended by a mechanical spring governed by Hooke's law [19]. The application of a bias voltage between the electrodes actuates them towards each other, but it is possible that the moving component to become unstable and collapse (pulls-in) onto the other [13, 16]. Residual stress and fringing field effects have also been shown to have great influence on the behavior of micro switches and strongly influence their failure characteristics [17, 18]. Recently, using the measured optical response and surface roughness topography as input [25]; realistic calculations have been performed to account for Casmir and electrostatic forces on actuation dynamics of micro switches. It was found that surface roughness ensured that stable equilibrium can be reached more easily than in the case of flat surfaces [25], stimulating further understanding of MEMS stability issues operating at separations $\leq 100$ nm.

So far, however, a detailed study of the sensitivity of actuation dynamics on a systematic variation of the measured optical properties of interacting materials without variation of their composition is missing. This motivated our attempts to explore MEMS actuation dynamics with phase change materials (PCMs) [8, 9], which are renowned for their use as active media in rewritable optical data storage (e.g., CD, DVD and Blu-Ray Disks), and their optical properties can be changed reversibly in response to a simple stimulus (e.g. local heating by a laser) leading to reversible switching between amorphous and crystalline phases. This is stimulated by the fact that we have already demonstrated that PCMs are promising to achieve significant force contrast ~25 % [8, 9] without composition changes, paving the way for a high repetition rate switchable force device with possible applications in MEMS.

## II. Force theory and modelling

As Fig. 1(a) illustrates, we consider for our study a moving sphere coated with gold (Au) interacting with a fixed plate coated with thick PCM film (optically bulk; thickness ≥100 nm) with optical properties, as depicted in Fig. 1(b) those of AIST ($Ag_5In_5Sb_{60}Te_{30}$) [8, 9]. The force in the sphere-plate geometry (widely used in force measurements by atomic force microscopy-AFM and MEMS [4-10]) is given by $F_C(z) = 2\pi R E_C(z)$ with R the sphere radius, z the sphere-plate separation (assuming z<<R), and $E_C(z)$ the Casmir energy in the parallel plate configuration that is calculated via Lifshitz theory [3]. Therefore, we have [3]

$$F_C(z) = k_b T R \sum\nolimits_{n=0}^{\prime} \sum\nolimits_{v=s,p} \int_0^\infty q \ln(1 - r_1^v r_2^v e^{-2|k_0|z}) dq \qquad (1)$$

The sum over the v-index is for the transverse electric and magnetic field modes(s=TE, p=TM). $r_i^{s,p}$(i=1, 2) are the Fresnel reflection coefficients $r_i^s = (k_0 - k_i)/(k_0 + k_i)$ and $r_i^p = (\varepsilon_i k_0 - \varepsilon_0 k_i)/(\varepsilon_i k_0 + \varepsilon_0 k_i)$ and $k_m = \sqrt{\varepsilon_m(\omega)(\omega^2/c^2) - q^2}$ (m=0, 1,2) with q an in-plane wave vector. The summation over the n-index is over the Matsubara frequencies $\omega_n = j(2\pi n k T)/\hbar$ with n=0,1,2…, where the n=0 term is taken only half [3]. The $\varepsilon_{1(\equiv Au)}(\omega)$ are the measured dielectric dataof Au [11], and $\varepsilon_{2(\equiv AIST)}(\omega)$ those of AIST [8,9] ($\varepsilon_0(\omega) = 1$ since vacuum is the intervening medium). It should be mentioned also that in the sphere-plane geometry we consider here the residual contact potential dependence on separation can lead to undesirable effects. Measurements of the contact potential difference vs. separation for PCMs [8] indicated a residual electrostatic contribution 1-6 % for 50 - 150 nm. Therefore, for a force contrast of ~25 % [8] this is not expected to play significant role, however one has to remain cautious concerning electrostatics.

The inset of Fig.1(b) shows calculations of $F_C(z)$ for both PCM states. Tests with three different optical data sets of Au [11] show also that the preparation conditions of the passive actuating material is playing minor role on the force contrast. To achieve, however, the force contrast with PCMs the area that undergoes phase transformation must be larger than the effective interaction area $A_C \approx L_C^2$ of the Casimir force between sphere-plate. For a sphere-plate separation z (<<R) we have $L_C \approx \sqrt{(2\pi/3)Rz}$ [26]. Moreover, the minimum thickness of the PCM film must be larger than the skin depth $\delta$ in the IR range where the crystalline PCM has strong absorption due to free carriers and contribute ~50 % of the force contrast [9]. For AIST we obtain $\delta = c/\omega_p \sim 100$ nm with c is the velocity of light and $\omega_p$ the free carrier plasma frequency [9]. For a minimum thickness $d_{PCM} \approx \delta$ (~100 nm) the energy necessary for crystallization is given by $E_{A \to C} = (L_C^2 d_{PCM}) \int_{T_{RT}}^{T_{cr}} C_p(T) \, dT$ with $C_p$ the specific heat capacity, $T_{RT}=300$ K, and $T_{cr}=451$ K the crystallization temperature for AIST [27, 28]. The energy necessary for amorphization (via melting at $T_{me}=807$ K) is given by $E_{C \to A} = (L_C^2 d_{PCM}) \int_{T_{RT}}^{T_{me}} C_p(T) \, dT + L_v$ where $L_v$ is the latent heat released during the phase transformation. Using the data of [27, 28], separation z=50 nm, which is comparable to the minimum separation in the force measurements with R=10.1 µm [8], we obtained respectively $E_{A \to C} = 38$ pJ and $E_{C \to A} = 337$ pJ. These energies are much larger than the work performed by the Casimir force over the closed path A→C→A, $\oint F_C(z) \, dz \sim 10^{-18}$ J, limiting the possibility to tap energy from vacuum fluctuations as the sole source to drive actuation.

Furthermore, the motion of the MEMS in Fig. 1(a) is described by the second law of Newton (assuming an initial impulse to trigger continuous actuation from an initial separation $L_o$), where the elastic restoring force $F_k = -K(L_0 - z)$ of a spring with stiffness K (Hooke's law) [19] counterbalances the attractive Casimir force $F_C(z)$ [22, 24, 25]:

$$m\frac{d^2z}{dt^2} = -K(L_0 - z) + k_B T R \sum_{n=0}^{\prime} \sum_{\nu=s,p} \int_0^\infty q \ln(1 - r_1^\nu r_2^\nu e^{-2|k_0|z}) dq \qquad (2)$$

For operation in air an additional dissipative hydrodynamic force has to be taken into account [29]. Here we will consider motion in vacuum and ignore any dissipation via the support base of the actuating element (a high quality factor system $Q>10^4$ [30]). Moreover, we consider actuators with resonance frequency $\omega = 300$ kHz (typical for AFM cantilevers and other MEMS) [31].

### III. Results and discussion

In order to obtain the equilibrium points of motion from Eq. (2) we define the bifurcation parameter $\lambda = F_{C(A)}(L_0)/KL_0$ [14, 22, 24], which is the ratio of the minimal Casimir force (in the amorphous PCM) and the maximal elastic restoring force, representing the relative importance of one force competing to the other. The locus of equilibrium points 'x' is obtained from Eq. (2) if we set $F_T = -K(L_0 - z^*) + F_C(x) = 0$ [22, 24, 25]. Solution yields for the bifurcation parameter $\lambda$

$$\lambda = (F_{C(A)}(L_0)/F_C(z^*))(1 - z^*/L_0). \qquad (3)$$

The critical equilibrium points where stiction occurs are characterized also by the condition $dF_T/dz^* (= K + dF_C/dz^*) = 0$ [22, 24, 25]. The dependence of the parameter $\lambda$ on the locus of equilibrium points $z^*$ is shown in Fig. 2(a) for both PCM states, and in comparison to the Au-Au system that is widely used in Casimir force measurements [4-10]. It is evident that the bifurcation parameter $\lambda$ is sensitive to phase transitions of the PCM from the amorphous to crystalline state.

The maximum of the bifurcation parameter $\lambda_{max}$ is approximately on the same location for both curves at $z^*_{max} \approx 0.75 L_o$. This is because the Casimir force shows an average power-law scaling $F_C(x) \sim z^{-p}$ with $p \approx 2.4$-$2.6$ (for $z < 200$ nm) [32], and it is known that if a force field scales as $\sim z^{-p}$ then the position of the maximum bifurcation parameter is given by $z^*_{max} \approx p/(1+p)L_o$ [21]. Moreover, as Fig. 2(a) shows if the spring constant is strong enough so that $\lambda < \lambda_{max}$, there will be two equilibria: the stationary point closest to $L_o$ is a stable center around which periodic solutions exist. In Fig. 2(a) the locus of points for $z^* > z^*_{max}$ corresponds to stable actuation leading to closed orbits (see Figs. 2(b) and 3(c)). However, if the spring constant is sufficiently weak so that $\lambda = \lambda_{max}$ then in this case there is only a single equilibrium, known as a center-saddle point [33], which is always unstable. For an even weaker spring constant K so that $\lambda > \lambda_{max}$ the motion is unstable, which is an example of a saddle-node bifurcation [33].

The solutions of Eq.(2) can be further investigated with the so-called phase portraits [33], which are plots of the velocity $dz/dt$ of the actuating element vs. the displacement z. In this case we can have: i) a periodic orbit around an equilibrium point and ii) unstable open orbit for a saddle point leading to pull-in instability or stiction (Figs.2(a), 4(b)). The existence of periodic solutions indicates that the spring is strong enough to counterbalance the attractive surface forces and prevent stiction. In this case the stable center around which periodic solutions exist will be accompanied by an unstable saddle-point equilibrium [33]. This is clearly manifested in Fig. 2(b) where closed orbits are shown for both amorphous and crystalline PCM, and soft spring constants $K \sim 10^{-3}$-$10^{-4}$ N/m which are used in ultra-sensitive systems [31]. The difference between the amorphous and crystalline phases is amplified for the part of the orbit that comes in close proximity to the plate where the Casimir force is the strongest, thus enhancing the possibility for stiction to occur. This will happen if the spring constant weakens further (entering the regime of the saddle-node bifurcation) and stiction can no longer be avoided (Fig. 2(c)). For relatively large separations, $z/L_0 > 0.6$, the actuating component follows a stable movement, which changes to a rapid pull-in for $z < 0.6L_0$, having high sensitivity to small changes of the spring stiffness K.

Another parameter that can be varied independently of K is the radius R of the sphere. Figs. 3(a) and 3(b) are contour plots of the bifurcation parameter $\lambda \propto 1/K$ for the crystalline and amorphous phases, respectively. Since $F_C(z) \propto R$ the magnitude of the Casimir force can be modulated with the size of the sphere. It is clear that increasing the radius enhances the sensitivity to the phase transition, but the tradeoff is that this requires higher values of K to prevent stiction. An example of solutions for a sufficiently large value of K is shown in Fig.3(c).

It should be noted that for a real system where the PCM undergoes a phase change from the amorphous to crystalline states a volume compression corresponding to ~5-8 % of the PCM film thickness can occur [34], which translates to additional force changes with surface separation. As a result proper feedback is required to readjust the actuating components at the same separation as performed during the dynamic force measurements in [8,9]. Moreover, explicit time dependence (e.g. forcing of the oscillations) would change the dynamics of the system considerably, because it would introduce an independent variable in phase space. The phase portraits and bifurcation diagrams presented in this manuscript would no longer be valid. Such a generalization would involve a different kind of mathematics, which is beyond the scope of this manuscript. However, qualitatively we can state that PCMs can be switched very fast ~$10^{-7}$-$10^{-8}$ sec [34], while typical actuation times for the phase maps shown here (Figs.(2)-(4)) are

~$10^{-6}$ sec. Therefore, it is possible to alter the system dynamics efficiently by phase change transformations.

Finally, up to now we assumed absence of dissipation, or equivalently having a MEMS of high quality factor Q>$10^4$ (dissipation~1/Q). To investigate the effect of intrinsic dissipation we considered in Eq.(2) the presence of a dissipative term of the form $(m\omega/Q)(dz/dt)$ with $K = m\omega^2$ [31] yielding $m(d^2z/dt^2) + (m\omega/Q)(dz/dt) = -K(L_0 - z) + F_C(z)$. Figure 4(a) shows that for Q≥$10^4$ the motion approaches that of a stable orbit, while as the Q factor decreases (inset Fig. 4(a)) it attenuates drastically faster for the amorphous PCM state. However, it is intriguing to investigate what is happening if the spring constant is sufficiently weak (below a critical value $K_{cr}$) to drive the system to stiction (S) under conditions of strong dissipation favoring attenuated motion (AM) towards stable equilibrium. This is shown in Fig. 4(b) where, as Q decreases, the S↔AM transition occurs rather sharply. For crystalline PCM the S↔AM transition occurs for $Q_C$≈480 (K(C)$_{cr}$=9x$10^{-5}$ N/m), while for amorphous PCM at a lower $Q_A$≈380 (K(A)$_{cr}$=7.5x$10^{-5}$ N/M). The critical $Q_C$ factor where the S↔AM transition occurs (at separation $z_{cr}$ just before the Casimir force becomes stronger than the elastic force) is given by Eq.(1) if we consider $m(d^2z/dt^2)≈0$ and $(m\omega/Q_C)|\dot{z}|≈|-K(L_0 - z_{cr}) + F_C(z_{cr})|$ with $|\dot{z}|≈\omega L_0$: substitution yields

$$Q_C = (m\omega^2 L_0)/|F_C(z_{cr}) - K(L_0 - z_{cr})|. \quad (4)$$

The latter yields for both PCM states $Q_{C(C)}$=498 and $Q_{C(A)}$=392, which compare well with the numerical results for $Q_{A,C}$ above.

## IV. Conclusions

In conclusion, the reversible amorphous to crystalline phase transitions in PCMs can have strong influence on nanoscale actuation of MEMS under the influence of Casimir forces. Although we considered here flat surfaces, this is justified for actuation separations >100 nm, since our former studies [8, 25] indicated morphology contributions only at lower separations. The phase portraits that characterize the actuation dynamics show strong sensitivity to changes of the Casimir force as the stiffness of actuating component decreases, which can also lead to stiction. On the other hand introduction of energy dissipation can prevent stiction by driving the system to attenuated motion towards equilibrium depending on the PCM state and system quality factor Q. Therefore, the use of PCMs that allow modulation of their optical properties can provide new schemes of actuation, where introduction of dissipation can prevent stiction of soft components.

## Acknowledgements

We would like to acknowledge useful discussions with J. Knoester, and support from the Zernike Institute of Advanced Materials, University of Groningen, The Netherlands.

# Figure captions

**Figure 1 (a)** Schematic of an actuated MEM system at initial separation $L_0 = 200$ nm with the acting forces. **(b)** Absorptive part $\text{Im}[\varepsilon(\omega)]$ of dielectric function vs. frequency $\omega$ for the optically active AIST film measured by ellipsometry for both the amorphous (A) and crystalline (C) phases. The inset shows the Casimir force vs. separation for both PCM phases.

**Figure 2 (a)** Bifurcation diagrams for the amorphous and crystalline phases of the AIST-Au system. For comparison we show also the bifurcation diagram of the Au-Au system. For illustration purposes we indicate two possible solutions if $\lambda < \lambda_{max}$ for AIST (A). **(b)** Phase portraits for AIST (A, C) and different (relatively weak) spring constants K. For all calculations we used $L_0=200$ nm, which is a typical surface separation for nanoscale actuation. **(c)** Phase portraits for AIST (C) states for different spring constants K (N/m) as indicated, R=10.1 μm, where the transition from stable (closed orbits) to unstable motion (open orbit) is shown.

**Figure 3 (a)** Contour plots of the bifurcation parameter $\lambda$ (indicated by the color bar) for the crystalline phase. The radius and the spring constant are varied independently. **(b)** As in (a), for the amorphous phase. **(c)** Phase portraits for AIST (A, C) and two different sphere radii R that are used in Casimir force measurement systems [4-10].

**Figure 4 (a)** Phase portraits for crystalline and amorphous PCM states, $K=10^{-4}$ N/m, R=10.1 μm, and two distinct Q factors: Q=10000 (main plot), Q=300 (inset). The arrows indicate the direction of the vector field. **(b)** Phase portraits of AIST (C) for different Q factors at a critical spring constant as indicated, R=10.1 μm, where the S↔AM transition (at $z_{cr}=117.4$ nm) occurs. The behavior for amorphous AIST is similar. The arrows indicate the direction of the vector field.

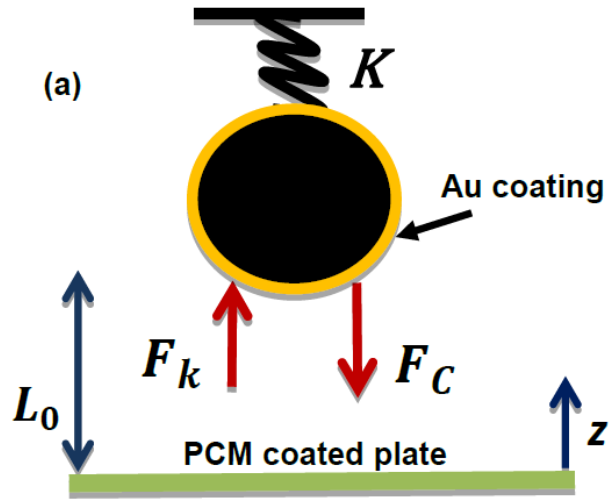

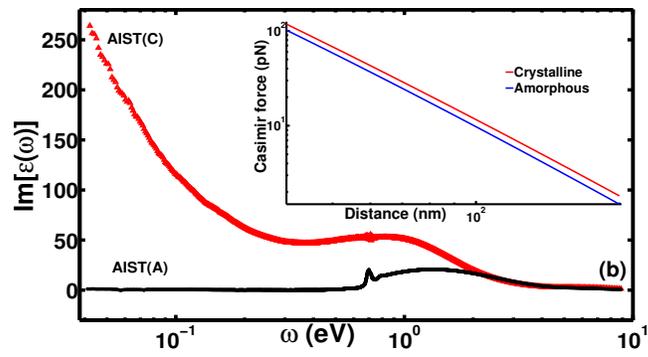

**Figure 1**

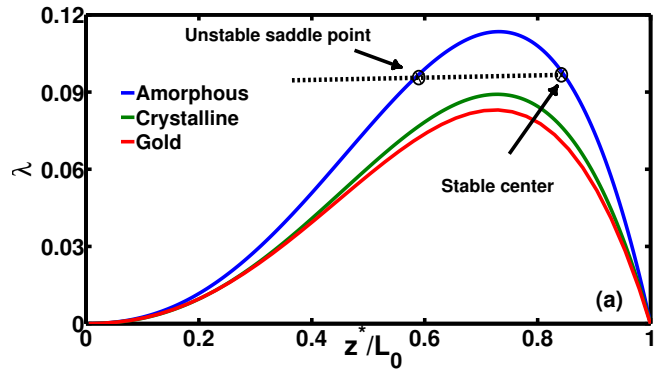

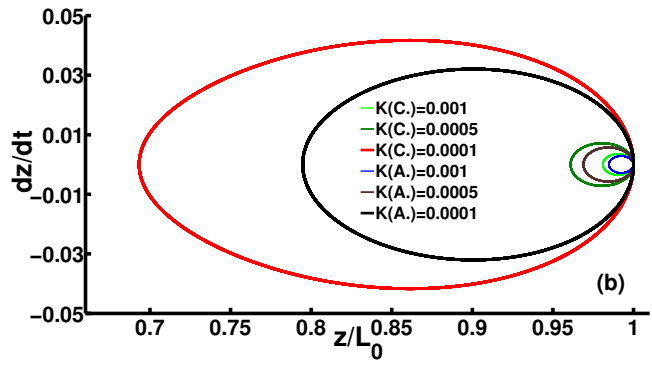

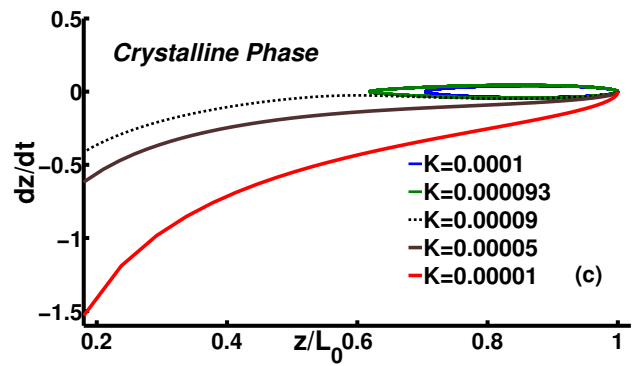

**Figure 2**

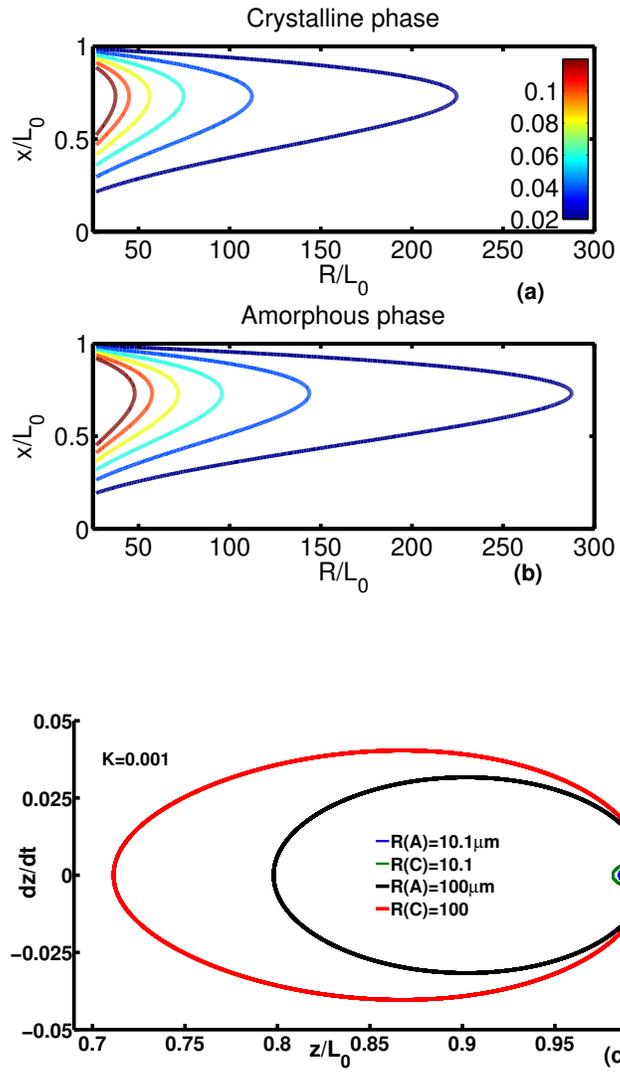

**Figure 3**

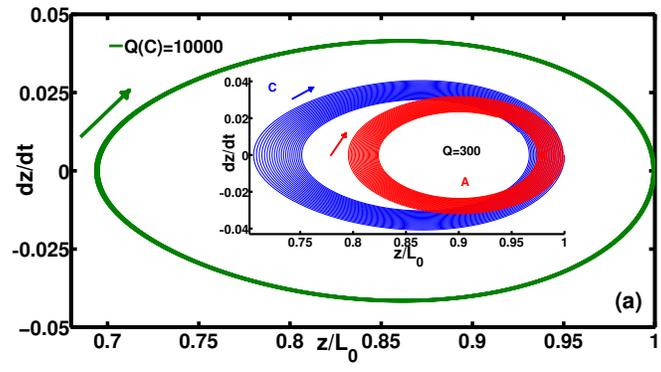

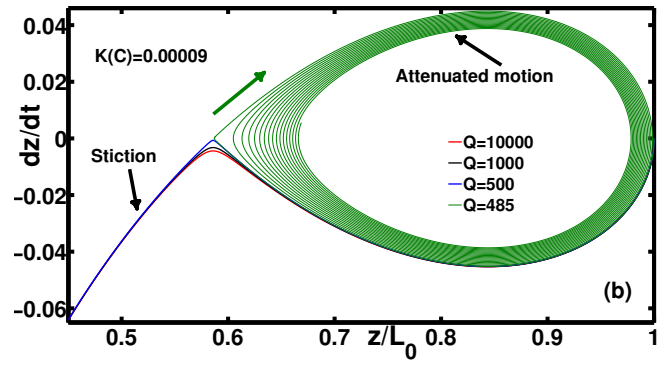

**Figure 4**